\documentclass[useAMS,usenatbib]{mn2e}
\usepackage{graphicx}
\title[X-ray and radio observations of G78.2+2.1]
	{X-ray and Radio Observations of the $\gamma$ Cygni Supernova Remnant G78.2+2.1}
\author[D.A. Leahy et al.]{ D.A. Leahy$^{1}$
\thanks{E-mail: leahy@ucalgary.ca}, K.~Green$^{1}$, 
S. Ranasinghe$^{1}$\\
$^{1}$Department of Physics $\&$ Astronomy, University of Calgary, Calgary,
Alberta T2N 1N4, Canada\\} 
\pagerange{\pageref{firstpage}--\pageref{lastpage}} \pubyear{2002}

\def\LaTeX{L\kern-.36em\raise.3ex\hbox{a}\kern-.15em
    T\kern-.1667em\lower.7ex\hbox{E}\kern-.125emX}

\begin{document}
\date{Accepted ; Received; in original form }
\pagerange{\pageref{firstpage}--\pageref{lastpage}} \pubyear{}

\maketitle

\label{firstpage}

\begin{abstract}

We analyse \textit{ROSAT} and \textit{Chandra} ACIS X-ray observations  and  HI absorption spectra of the 
 $\gamma$ Cygni supernova remnant (G78.2+2.1, DR4).
The \textit{ROSAT} All-Sky-Survey image shows G78.2+2.1 has 
an adjacent limb-brightened shell north of it.
A new \textit{ROSAT} mosaic shows details of the X-ray emission over 
the entire face of G78.2+2.1.
We also create \textit{Chandra} mosaics which cover much of
the northern rim and central regions of G78.2+2.1. 
HI absorption spectra result in association of G78.2+2.1 with the $\gamma$ Cygni nebula, with
distance 1.7 to 2.6 kpc. 
Chandra spectra for G78.2+2.1 give an X-ray temperature of 0.6-1.2 keV (90 \% error), 
and that a Sedov model has age of 6800-10000 yr. 
A compact power-law X-ray source in G78.2+2.1 
is consistent with the same distance as G78.2+2.1.
The northern X-ray shell is identified with a B3 star at distance of 980 pc and is proposed as 
a stellar wind bubble.

\end{abstract}


\section{Introduction}

G78.2+2.1 (or DR4) is a nearby shell-type supernova remnant (SNR) 
with a radio diameter of $\sim$60 arcmin (Higgs, Landecker, \& Roger, 1977). 
G78.2+2.1 has been of particular interest recently because
of its possible association with unidentified $\gamma$-ray sources in the Galactic
plane. If it is associated, then the question of whether the $\gamma$-rays are produced
by a young pulsar or by high-energy particles accelerated by the supernova shock
can be addressed.  

At a Galactic longitude of 78$^\circ$, G78.2+2.1 is located in the Cygnus X region of
the Galactic plane, about 2.5 degrees from the center of Cygnus X. 
Cygnus X is about 6 degrees in diameter and is 
one of the most nearby massive star forming regions within our
Galaxy (for a review e.g. see Kn{\"o}dlseder, 2004).
The most massive stars in Cygnus X reside at its center in the 
Cyg OB2 association, which
is at a distance of 1.3-1.5 kpc as determined recently by VLBI parallax
measurements (Rygl et al., 2012). The age of Cyg OB2 is 3-4 Myr, with total Lyman
continuum luminosity of $\sim 10^{51}$ ph/s and 
total mechanical luminosity of a few $10^{39}$ erg/s (Kn¨odlseder, et al. 2002)
The Cyg OB2 association injects enough kinetic energy, that a superbubble shell
should have been created with radius of $\sim$60pc and velocity 20 km/s
(Lozinskaya et al., 2002). X-rays from the superbubble interior were detected
by Cash et. al (1980), but other evidence is confusing.
A recent study of molecular gas, using the CO line, across a significant area 
($\sim$2 degrees by 4 degrees) of Cygnus X was carried out by  Gottschalk et al.(2012). 
In summary, their findings are: 
i) there are three layers of gas across the region, 
at radial velocities of 0 to 8 km/s (layer 1),
-10 to 0 km/s (layer 2) and above 8 km/s (layer 3).
ii) Layer 1 shows extensive HI self-absorption, is nearest to us and is associated
with cold gas and dust in the Great Cygnus Rift, 
located at a distance of 500-800 pc (Schneider et al. 2006).
HI absorption profiles indicate that layer 3, associated with W75N and DR17, 
is in front of layer 2, associated with DR21, G82.6+0.4 and IRAS 23350+4126.  
iii) The fact that layer 3 is redshifted with respect to the expected radial velocity 
of Cyg OB2 and layer 2 is blueshifted is not consistent with the picture of an
expanding shell of molecular gas around OB2.  
An important implication of this study is that radial velocities do not fit in with
a simple circular rotation model for the Galaxy in the Cygnus X region.

The SNR G78.2+2.1 has been observed several times at radio, 
optical and X-ray wavelengths.
In radio, several spectral index studies
have been carried out, with one of the most comprehensive by Zhang et al. (1997).
Optical and X-ray observations were reported by Lozinskaya et al. (2000).
They searched for but did not find an optical shell associated with G78.2+2.1.
The \textit{ROSAT} X-ray image was found to have a shell structure similar to the
radio image. The ASCA X-ray spectra were analysed to find 
temperatures of 1.4$\pm$0.1 keV  for an equilibrium ionization plasma model
and 2.8$\pm$0.4 keV for a non-equilibrium ionization plasma model.
Uchiyama et al. (2002) also analysed ASCA X-ray observations of G78.2+2.1
and showed images in 3 energy bands: 0.7-1 keV, 1-3 keV and 4-7 keV.
The two lower energy band images show an irregular shell structure for G78.2+2.1
and the high energy band shows two compact bright spots, labelled C1 and C2.
The compact spots are not coincident with the EGRET $\gamma$-ray sources
2EG J2020+4026 or 3EG J2020+4017.
Their ASCA spectral analysis shows that the diffuse emission (their regions R1, R2 and R3) is well fit with
an 0.5-0.8 keV thermal plasma, and that their regions C1 and C2 have an additional
power-law components with spectral indices $\sim 0.8-1.5$. 

Mavromatakis (2003) imaged the G78.2+2.1 region in H$\alpha$+[NII], [SII] and [OIII] filters,
noting that the region is dominated by emission from several HII regions.
Long-slit spectra at one position detected shock-heated gas (with [SII]/H$\alpha$ $\simeq$
0.6), with estimated shock velocity below 100 km/s and pre-shock density $\sim$20 cm$^{-3}$.
The H$\alpha$/H$\beta$ ratio was 10 for this region, equivalent to a column density
of 6$\times$10$^{21}$cm$^{-2}$.
Bykov et al. (2004) report a bright hard X-ray (25-40 keV) clump which is nearly coincident with the ASCA C2 source, with a flux of  1.7$\times$10$^{-11}$erg cm$^{-2}$s$^{-1}$.
Ladouceur \& Pineault (2008) use the Canadian Galactic Plane Survey continuum data for G78.2+2.1 
to better delineate the extent of the SNR and the spatial spectral index variations, finding an
spatially-averaged spectral index of 0.75$\pm$0.3, steeper than previous studies.
They also associate HI structures in two different velocity ranges with the SNR.

Trepl et al. (2010) investigate the 3.77 Hz $\gamma$-ray pulsar PSR J2021+4026, which is located 7.8 arcmin
from the center of G78.2+2.1. It has a spin-down age of 77 kyr and is associated with an XMM/Newton
X-ray source 2XMMJ202131.0+402645 and Chandra X-ray source S21. S21 has the best determined position
of 
($\alpha$(J2000)=$\mathrm{20^h21^m30.553^s}$,  $\delta$(J2000)=$40^{\circ}26^{\prime}46.89^{\prime\prime}$)
%RA (J2000) = 20h21m30.553s, Dec. (J2000) = +40deg26'46.89" 
with a position error of $\sim 1^{\prime\prime}$.
The X-ray spectrum is best fit by a blackbody, although bremmstrahlung and MEKAL models cannot be ruled 
out. It also has quite uncertain 
X-ray column density and thus distance: 2(+5,-2) or 6(+6,-4)$\times$10$^{21}$cm$^{-2}$, 
respectively for black-body
and bremsstrahlung emission models, and higher values of 10(+8,-5) or 
16(+6,-3)$\times$10$^{21}$cm$^{-2}$ for power-law and MEKAL models. Thus the association
with the SNR G78.2+2.1 is uncertain. 
Wienstein et al. (2011) report a TeV source in the direction of G78.2+2.1, labelled VER J2019+407.
It was a 7.5 sigma detection (268$\pm$34 net counts), enough to get a position but not a spectrum:
($\alpha$(J2000)=$\mathrm{20^h20^m0^s}$,  $\delta$(J2000)=$40^{\circ}49^{\prime}12^{\prime\prime}$) 
with 7$^{\prime}$  radius 95\% confidence error circle.

In this paper, our focus is to understand the physical properties of G78.2+2.1. 
We carry out a comprehensive analysis of X-ray observations from \textit{ROSAT} and \textit{Chandra}
observations, to delineate the X-ray structure of G78.2+2.1 and to sort out the conflicting
results of previous X-ray spectral studies on G78.2+2.1. 
Also, we obtain new HI absorption spectra by analysis of the HI line and 1420 MHz continuum observations 
from the Canadian Galactic Plane Survey (CGPS, English et al. 1998, Taylor et al. 2003). 
Section 2 describes the 
observations and data analysis, section 3 presents the analysis and
results, and section 4 presents conclusions.

\section{Observations and Data Analysis}

\subsection{X-ray Data}

The \textit{ROSAT} all-sky survey (RASS) covers all of G78.2+2.1 and its surroundings. 
We found that \textit{ROSAT} PSPC pointed observations covered all of G78.2+2.1 but not its surroundings,
and that \textit{Chandra} observations covered only part of G78.2+2.1. Thus we obtained all three sets of
data for analysis.
The archival X-ray observations from \textit{ROSAT} and \textit{Chandra} for the G78.2+2.1 region 
were obtained from NASA's HEASARC website.
Data (images, photon event lists and associated files) which covered any part of G78.2+2.1 or its
nearby surroundings were obtained. 

The list of the \textit{ROSAT} pointed observations, with pointing centers and net exposure times, 
is given in Table 1.
The \textit{ROSAT} data were processed using the ftools and HEASOFT software packages, available at the HEASARC website. 
ROSAT Guest Observer Facility analysis procedures were used (OGIP Memo 94-010).
The following ftools were used to make the ROSAT mosaic: 
xselect (to limit the range of pha channels used in each image), 
fimconcat (to make a blank image), 
fimgmerge (to add in the individual counts images)
(a separate image was made for the counts map and for the exposure map), 
farith (to create a count-rate map from counts and exposure maps), 
and fgauss (to smooth the count-rate map). 
To make an image using only the central high 
resolution parts of each PSPC image, we used xselect on the individual counts and exposure images before
merging.

The list of \textit{Chandra} observations is given in Table 2.
The standard archive ACIS good time intervals were used as no significant flares were seen in the data.
The \textit{Chandra} data were processed and spectra extracted using the CIAO  and \textit{Chandra} threads from the \textit{Chandra} Science
Center (http://cxc.harvard.edu). The \textit{Chandra} individual soft and hard band images were created using the CIAO (Version 4.4) package, and combined using the ftools software package. CIAO version
4.4 and CALDB version 4.4.8 were used. %oops should be CIAO version 4.4. not 4.4.8
Similar to the methods for creating the ROSAT mosaics, a combined counts image and a combined exposure image were created, then divided to obtain the count-rate map, and then Gaussian smoothing was applied.

\subsection{Radio Data}

Radio observations in 1420 MHz continuum and the HI spectral line were made as part 
of the Canadian Galactic Plane Survey (CGPS, English et al., 1998, Taylor et al., 2003). 
We extracted the 1420 MHz image of G78.2+2.1 from the survey data. 
We used the HI line data together with the continuum data to
determine HI absorption spectra for G78.2+2.1.
Since it is an extended source rather than a point source,
the standard formula for HI absorption spectra does not apply. 
The background region has to be chosen to be near the
continuum peak, in order to minimize the difference in the
Galactic HI distribution along the lines-of-sight to source
and background regions (Leahy \& Tian, 2008, and Tian \&
Leahy, 2008; see Leahy \& Tian, 2010 for a review). 
Because the brightest continuum emission and adjacent
background emission were separated by a curved boundary,
instead of boxes for source and background we used regions
defined by T$_B$. We used the software program 'meanlev'
in the DRAO export package which allows one to extract
spectra for source and background regions defined by user
specified T$_B$ levels.

\section{Results}

\subsection{HI Absorption Spectra}

We describe the radio image and absorption spectra first, 
because we use the image to locate the extent of
the supernova remnant on the X-ray images.
Fig.1 shows the 1420 MHz continuum image in the area around
G78.2+2.1, in Galactic coordinates (Ladoucer and Pineault, 2008, published a similar image). 
The image shows the shell-like radio morphology, with brightest filaments at the bottom (low b), 
and very faint parts of the shell at upper right and at lower left. 
As noted previously (e.g. Higgs et al., 1977), the $\gamma$ Cygni nebula (Drake, 1959) is located
at l,b=78.18$^\circ$,+1.82$^\circ$ and is a small bright region on the face of G78.2+2.1.
The $\gamma$Cygni nebula has a thermal radio spectrum;
the other regions have a non-thermal spectrum.

HI absorption spectra were constructed for a number (7) of different continuum emission areas
across G78.2+2.1, including  the $\gamma$ Cygni nebula. 
The Galactic coordinates (l,b) of the centers of these regions
(which are extended unless noted) are:
%are listed in Table 1.
%(need a list of coords of centers of the 9 regions:
(1)(point source) 78.51$^\circ$,+2.32$^\circ$ ; (2)(point source) 77.83$^\circ$,+2.40$^\circ$;
(3) 78.29$^\circ$,+2.46$^\circ$; (4)($\gamma$ Cygni nebula) 78.18$^\circ$,+1.82$^\circ$;
(5) 78.01$^\circ$,+1.78$^\circ$; %(6) 78.08,+1.73; omit 6, replace by number 8
(6) 78.37$^\circ$,+1.75$^\circ$; and (7) 78.11$^\circ$,+1.73$^\circ$. %and 
%We next summarize the results of the HI absorption spectra.
The emission spectra for all 7 regions are similar to the spectra shown in 
Figure 2 for region 7 of the SNR and for  the $\gamma$ Cygni nebula.
Region 7 is the bright filament just below the $\gamma$ Cygni nebula (region 4).  
The spectrum contains emission from HI interior to the solar circle at
positive radial velocities (0 to +25 km/s) and from the outer galaxy on the
far side of the solar circle in the direction of l=78.2$^\circ$ (0 to -100 km/s).

The absorption spectra for the 7 regions can be separated into two groups.
One group shows absorption spectra nearly identical to that of the bright filament of region 7.
It demonstrates absorption at all positive velocities and at negative velocities above -8 km/s.
The other regions in this group are 3, 4 (the $\gamma$ Cygni nebula), 5 and 6.
Region 3 is the brightest diffuse emission in the upper 
part of G78.2+2.1 (see Fig.1 left). Because it has low continuum 1420 MHz brightness, the 
HI spectrum is quite noisy but shows clear absorption at velocities more positive
than $\sim$-10 km/s. 
Region 5 is of lower continuum brightness than region 4 or 7 so has more 
noise but otherwise is very similar to the spectrum for region 7.
Region 6 is the filament at the lower left rim of G78.2+2.1 and also has an HI absorption spectrum very similar to region 7. Thus all of the diffuse emission regions 
in G78.2+2.1 were found to show absorption at velocities greater than -8 km/s but no
absorption for velocities less than this value.

The total Galactic column density in the direction of G78.2+2.1 is 
$N_H=1.25\times10^{22}$ cm$^{-2}$ from the 'nh' ftool, which is based on HI surveys. 
From an integration of the absorption spectrum of Fig. 1 (right) we find  
$N_H=6.2\times10^{21}$ cm$^{-2}$ for G78.2+2.1.
This value includes the correction for optical depth (see e.g. Binney 
\& Merrifield(1998) pp.468-473).

The two point sources make up the other group of absorption spectra. 
Region 1 is the point source at the upper left interior of G78.2+2.1  and 
region 2 is the  point source at the upper right. Both clearly shows 
absorption at all velocities (+25km/s to $<$-100 km/s), so are identified as
extragalactic background radio sources. 

\begin{figure*}	
	\includegraphics[scale=0.6, angle=270]{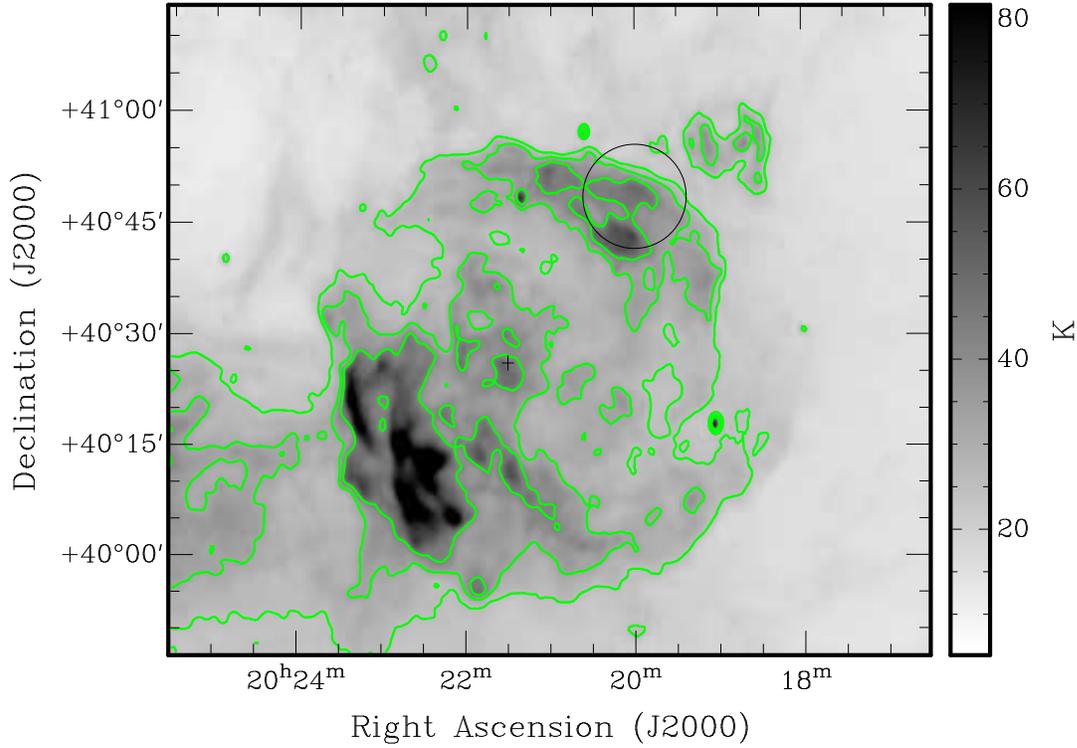}%{g78_1420_c253242.eps}
	\caption{1420 MHz image of G78.2+2.1, with contours at T$_B$= 25, 32 and 42 K. The black circle 
	indicates the error circle for VER J2019+407, the black plus indicates the position of 
	PSR J2021+4026.}
\end{figure*}

\begin{figure*}
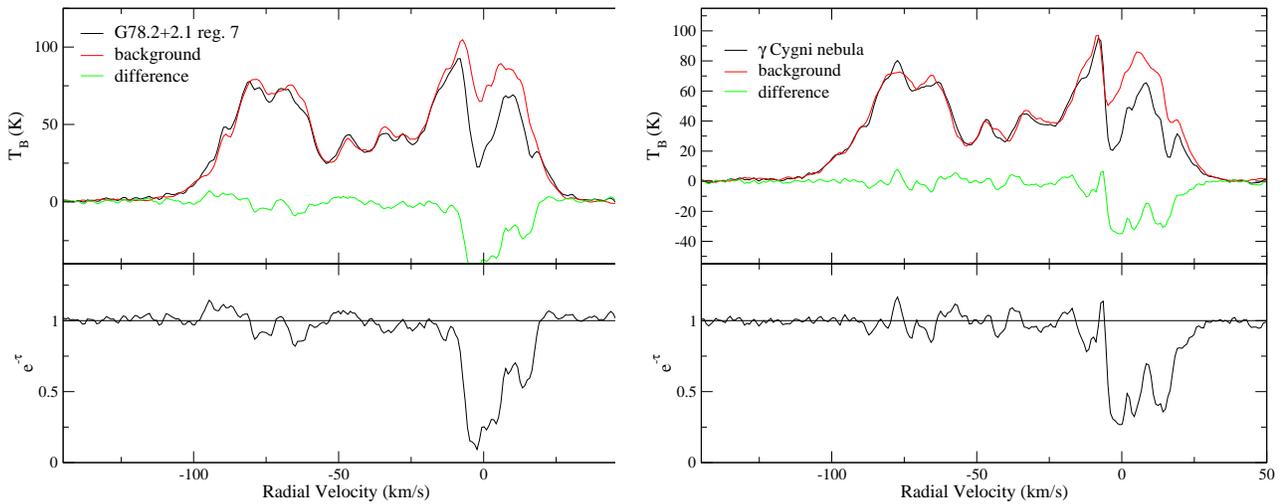
	
	\includegraphics[scale=0.35]{region7july2013.eps}
	\includegraphics[scale=0.35]{region4july2013.eps}
	\caption{(left) HI emission and absorption spectra for G78.2+2.1 region 7. The black line in the lower panel is the absorption spectrum. (right) HI emission and 
	absorption spectra for the $\gamma$ Cygni nebula.}
\end{figure*}

\subsection{X-ray Images}

\subsubsection{ROSAT}
In Figure 3(a), we present the flux-corrected \textit{ROSAT} all-sky image of the G78.2+2.1 region
in the 0.5-2.0 keV band. This image is overlaid with the 25, 32, and 42 K 1420 MHz 
contours of G78.2+2.1 (see Fig. 1) and includes the region north of G78.2+2.1. 
From this RASS image, we can see that there is significant X-ray emission associated with 
the $\gamma$Cygni region. There is an irregular X-ray shell that coincides with 
radio emission of the SNR G78.2+2.1. However, there is also emission extending north of the remnant. 
This extra emission roughly takes the form of a second shell centered near ($\alpha$(J2000)=$\mathrm{20^h20^m30^s}$, $\delta$(J2000)=$42^{\circ}30^{\prime}00^{\prime\prime}$).
The second shell does not show any evidence of radio emission: adjacent CGPS radio mosaics 
were examined and show that there is no radio shell associated with the x-ray emission. 

We combined the \textit{ROSAT} PSPC pointed observations into a single mosaic image in two different ways. 
First we included the
entire field of PSPC pointing in the mosaic. This results in maximum exposure time (and thus
sensitivity) but because of the rapidly increasing size of the point-spread-function with off-axis angle,
this also results in poor spatial resolution. Next we excluded the outer areas of each pointing
to exclude the low resolution data and made a higher resolution mosaic. Because of mirror vignetting,
the loss of sensitivity compared to the first mosaic is small.  The mosaics were made for
both soft (0.1-0.4 keV) and hard (0.5-2.0 keV) bands. 
The hard and soft maps were smoothed using a 2 arcmin Gaussian. 
The soft image shows only faint diffuse emission with no emission associated with G78.2+2.1.
This can be attributed to the X-ray absorption column density of the intervening interstellar 
medium, which prevents any emission from G78.2+2.1 from below 0.4 keV from reaching us.

The high-resolution \textit{ROSAT} PSPC 0.5-2.0 keV X-ray image is shown in Figure 3(b), overlaid with the
1420 MHz radio contours of G78.2+2.1.
It reveals that the X-ray emission forms an incomplete shell within the radio shell of the SNR. 
The brightest X-ray emission is located at the south-west radio boundary of the SNR.
In contrast, where the radio emission is brightest (in the south-east corner of G78.2+2.1, near 
%RA20h22.7m DEC40d15m) 
$\alpha$(J2000)=$\mathrm{20^h22.7^m}$, $\delta$(J2000)=$40^{\circ}15^{\prime}$
there are just several small spots of X-ray emission.  
Other prominent X-ray emission is seen inside the northern boundary of G78.2+2.1. 
This coincides with the second brightest area in radio emission (see Fig. 1). 
The X-ray and radio emission at the west boundary and in the western interior part of G78.2+2.1
are both very faint.
There is clear X-ray emission beyond the northern boundary of the radio SNR. 
The morphology of this emission, seen in Fig. 3(b), 
is fully consistent with that seen in the lower resolution RASS
image of Fig. 3(a), which shows the full extent of the large X-ray shell centered 
near ($\alpha$(J2000)=$\mathrm{20^h20^m30^s}$, $\delta$(J2000)=$42^{\circ}30^{\prime}00^{\prime\prime}$). 
From Fig. 3(a) and 3(b), it is not possible to tell whether the emission along the 
northern boundary of G78.2+2.1 originates from the SNR itself, or from the northern shell, 
or from a combination of both.

\begin{table}
\caption{\textit{ROSAT} PSPC pointed observations}
\label{symbols}
	\begin{tabular}{@{}lcccc}
	\hline
	Dataset ID  & RA($^\circ$)  & Dec($^\circ$) & live-time & Observation date\\
	\hline
	rp400386n00 & 304.790 & 40.350 & 8479 s & 1993-12-01\\
	rp500238n00 & 305.530 & 40.180 & 8155 s & 1993-10-23\\
	rp500244n00 & 305.050 & 40.690 & 7403 s & 1993-10-22\\
	rp500245n00 & 304.940 & 40.180 & 8668 s & 1993-10-26\\
	rp500246n00 & 305.630 & 40.580 & 5329 s & 1993-10-27\\
	rp900157n00 & 305.450 & 40.370 & 3864 s & 1991-11-24\\
	\hline
	\end{tabular}
\end{table}

\begin{figure*}	
	\includegraphics[scale=0.43, angle=270]{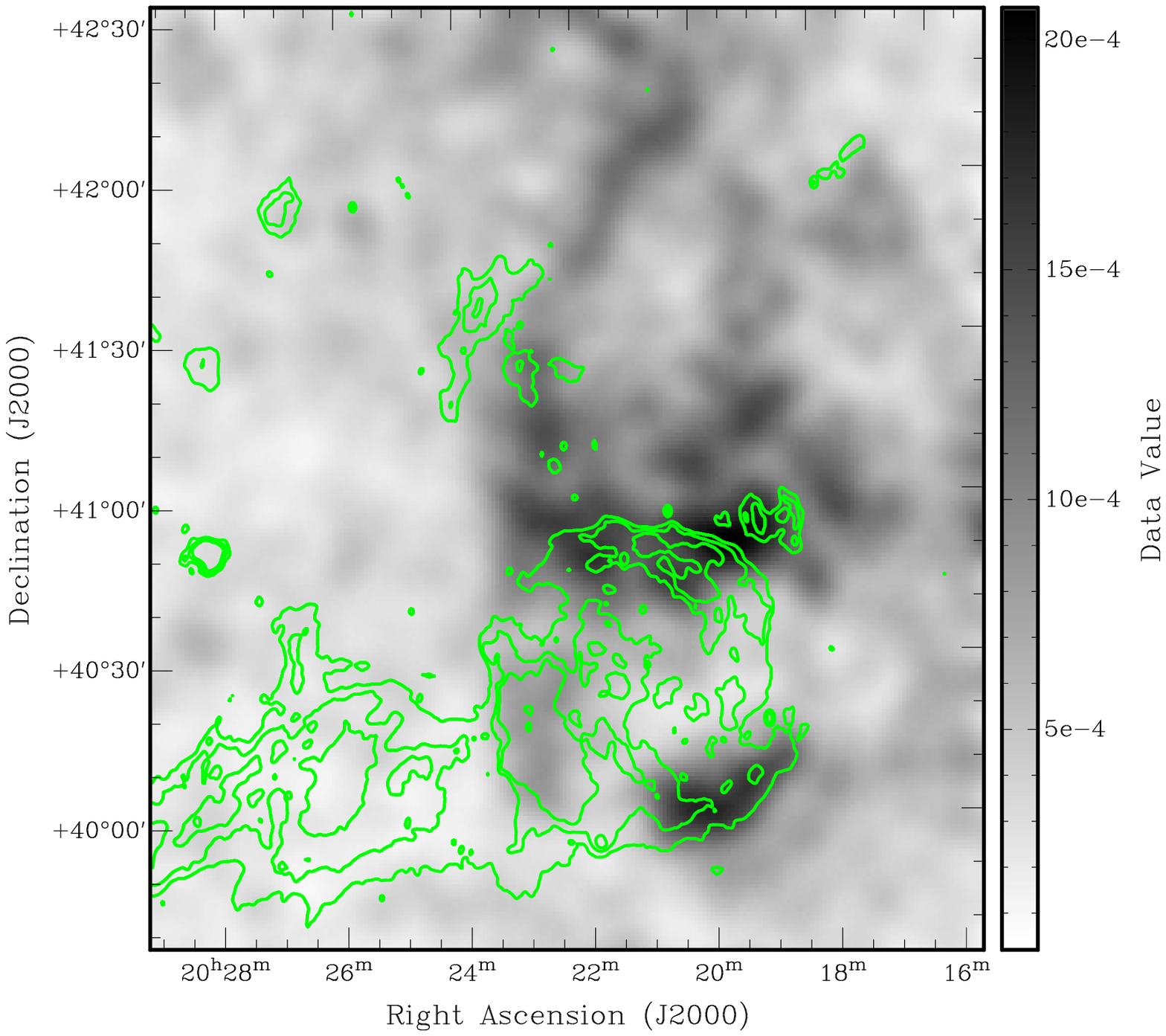}
	\includegraphics[scale=0.4, angle=270]{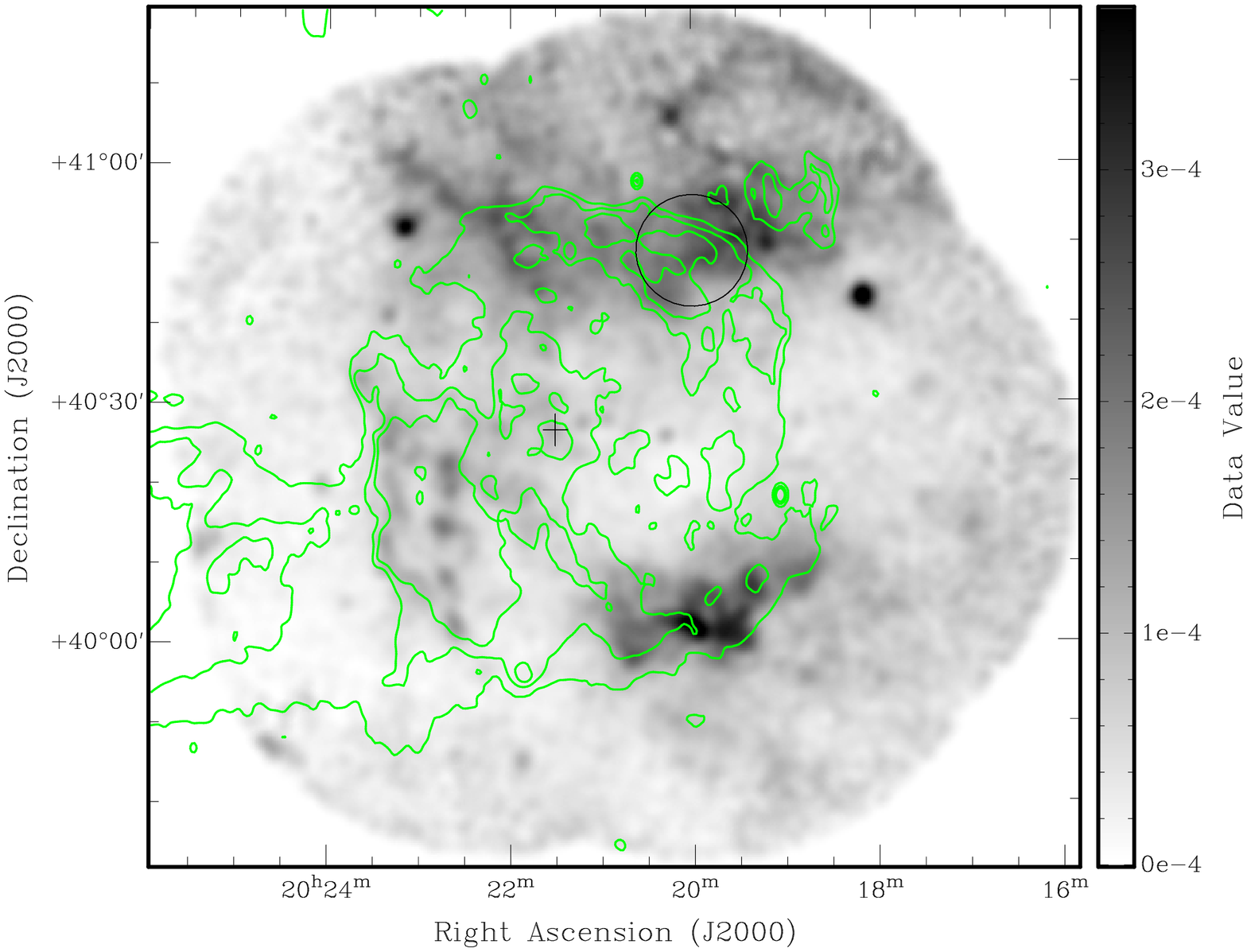}%{g78ctrcirccon253242.eps}
	\caption{(a) RASS survey image in counts per second. (b) 
	high resolution 0.5-2.0 keV \textit{ROSAT} PSPC mosaic smoothed with a 2' Gaussian function.
	The black circle 
	indicates the error circle for VER J2019+407, the black plus indicates the position of 
	PSR J2021+4026. 
	(a) and (b) are overlaid with the 1420 MHz radio contours at T$_B$= 25, 32 and 42 K.}
\end{figure*}

\subsubsection{\textit{Chandra}}
The SNR G78.2+2.1 has not been completely covered by \textit{Chandra} observations, however significant
parts of the north to north-west and central parts of G78.2+2.1 have been observed. 
Soft band (0.5-3.0 keV) and hard band (3.0-7.0 keV) image mosaics, with 4 arcsec pixels, were constructed using all
of the available \textit{Chandra} ACIS pointed observations of G78.2+2.1 (see Table 2).
The images were smoothed with a 12 arcsec Gaussian to better show the diffuse emission. 
Figure 4 shows the resulting 0.5-3.0 keV \textit{Chandra} image and Figure 5 shows the resulting 
3.0-7.0 keV \textit{Chandra} image. 
We included data from the I0, I1, I2, I3 and S2 ACIS chips, but we decided to 
not include data from the  backside illuminated spectroscopy chip (S3) because 
of its significantly different response to X-rays. 0.5-3 keV images made 
which include the S3 chip show significantly higher count rate over the area of the
S3 chip, and do not improve the visibility of diffuse emission in the region. 
The \textit{Chandra} images have considerably higher sensitivity and much higher spatial resolution than
the \textit{ROSAT} images, or any previous X-ray images of the $\gamma$ Cygni region. 
The \textit{Chandra} soft image (Fig. 4) shows that much of the X-ray emission is resolved into
point sources, but that there still exists significant diffuse emission in the region.
Point sources were identified using both wavdetect and the Chandra XASSIST 
catalog to verify the detection of point sources in the area. 
The diffuse emission is brightest along and interior to the northern boundary of G78.2+2.1, 
as delineated by the radio contours.
There is also significant diffuse emission in the center of G78.2+2.1 (coinciding with the brighter
radio emission which is left of the $T_B$=32K radio contour from 
($\alpha$(J2000)=$\mathrm{20^h20^m30^s}$, $\delta$(J2000)=$42^{\circ}07^{\prime}00^{\prime\prime}$) to ($\alpha$(J2000)=$\mathrm{20^h21^m00^s}$, $\delta$(J2000)=$40^{\circ}30^{\prime}00^{\prime\prime}$).
This diffuse emission can also be seen in the \textit{ROSAT} PSPC mosaic of Fig. 3(b).

Fig. 5 shows the hard band (3-7 keV) \textit{Chandra} image mosaic. 
This shows no diffuse emission, but does show a few small extended sources. 
There is an area of diffuse hard emission at ($\alpha$(J2000)=$\mathrm{20^h21^m24^s}$, $\delta$(J2000)=$40^{\circ}50^{\prime}00^{\prime\prime}$) that coincides with the hard source C1 described by Uchiyama et al. (2002). 
However, with the high resolution \textit{Chandra} image, we find that the 
 hard X-ray source located in the 6' region identified as source C2 by Uchiyama et al (2002)
 is a point source rather than an extended source. In the next section we consider
the X-ray spectral analysis for this point sources and the diffuse emission.

\begin{table}
\caption{\textit{Chandra} ACIS pointed observations}
\label{symbols}
	\begin{tabular}{@{}lcccc}
	\hline
	ObsID  & RA($^\circ$)  & Dec($^\circ$) & live-time & Observation date\\
	\hline
	2826 & 305.050 & 40.650 & 19960 s & 2002-07-09\\
	2827 & 305.350 & 40.810 & 9560 s & 2002-03-15\\
	3856 & 305.071 & 40.436 & 30360 s & 2003-01-26 \\
	5533 & 305.254 & 40.297 & 14350 s & 2005-02-06 \\
	12676 & 304.970 & 40.790 & 47360 s & 2011-01-08 \\
	\hline
	\end{tabular}
\end{table}

\begin{figure*}
 \includegraphics[scale=0.9, angle=0]{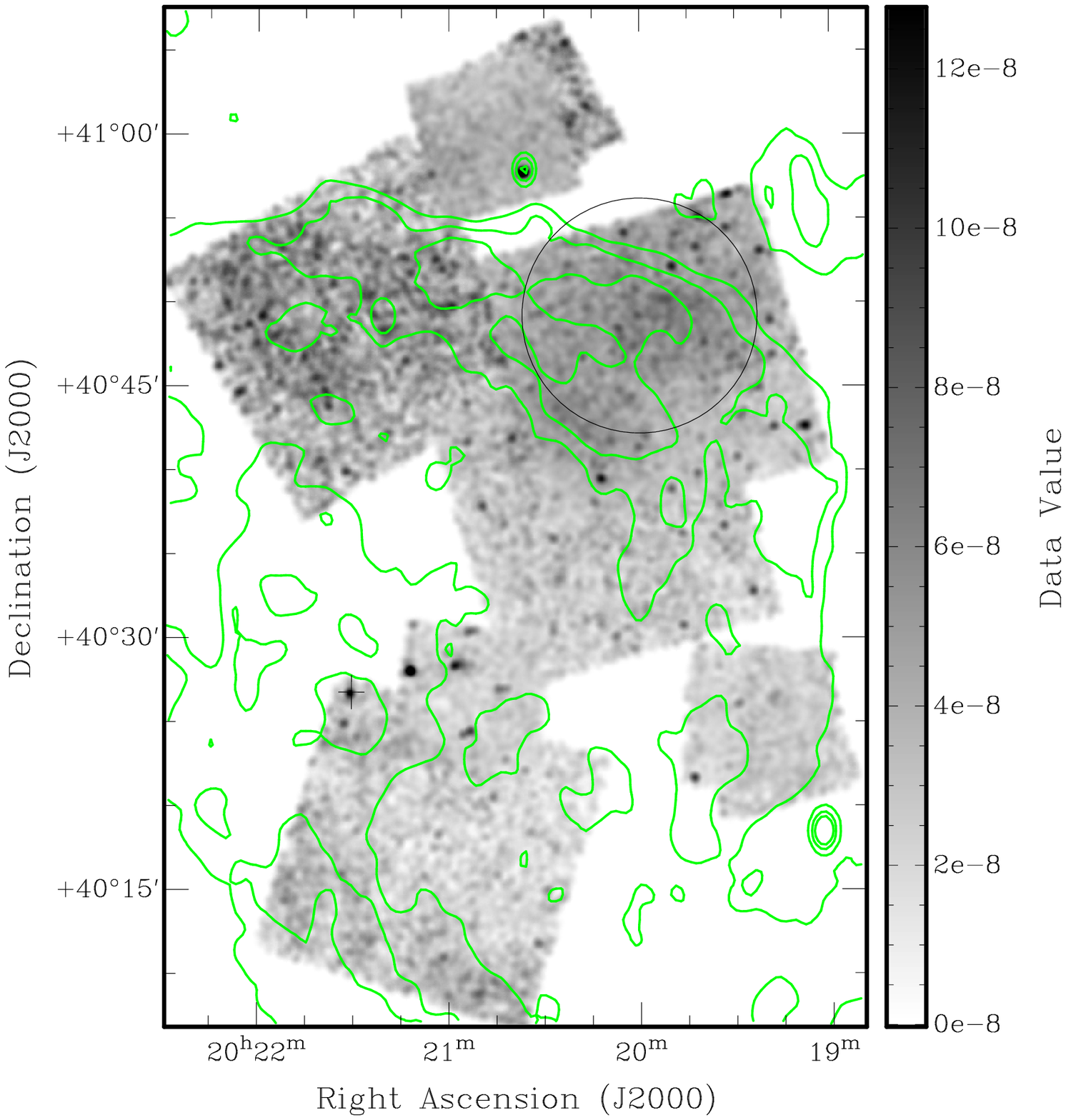}%{05-3_cts_ps.eps}
  \caption{ 0.5-3.0 keV \textit{Chandra} mosaic, overlaid with the 1420 MHz radio contours at T$_B$= 25, 32 and 42 K and smoothed with a 12 arcsecond Gaussian function. The black circle 
	indicates the error circle for VER J2019+407, the black plus indicates the position of 
	PSR J2021+4026.}
 \end{figure*}

 \begin{figure*}
 \includegraphics[scale=0.9, angle=0]{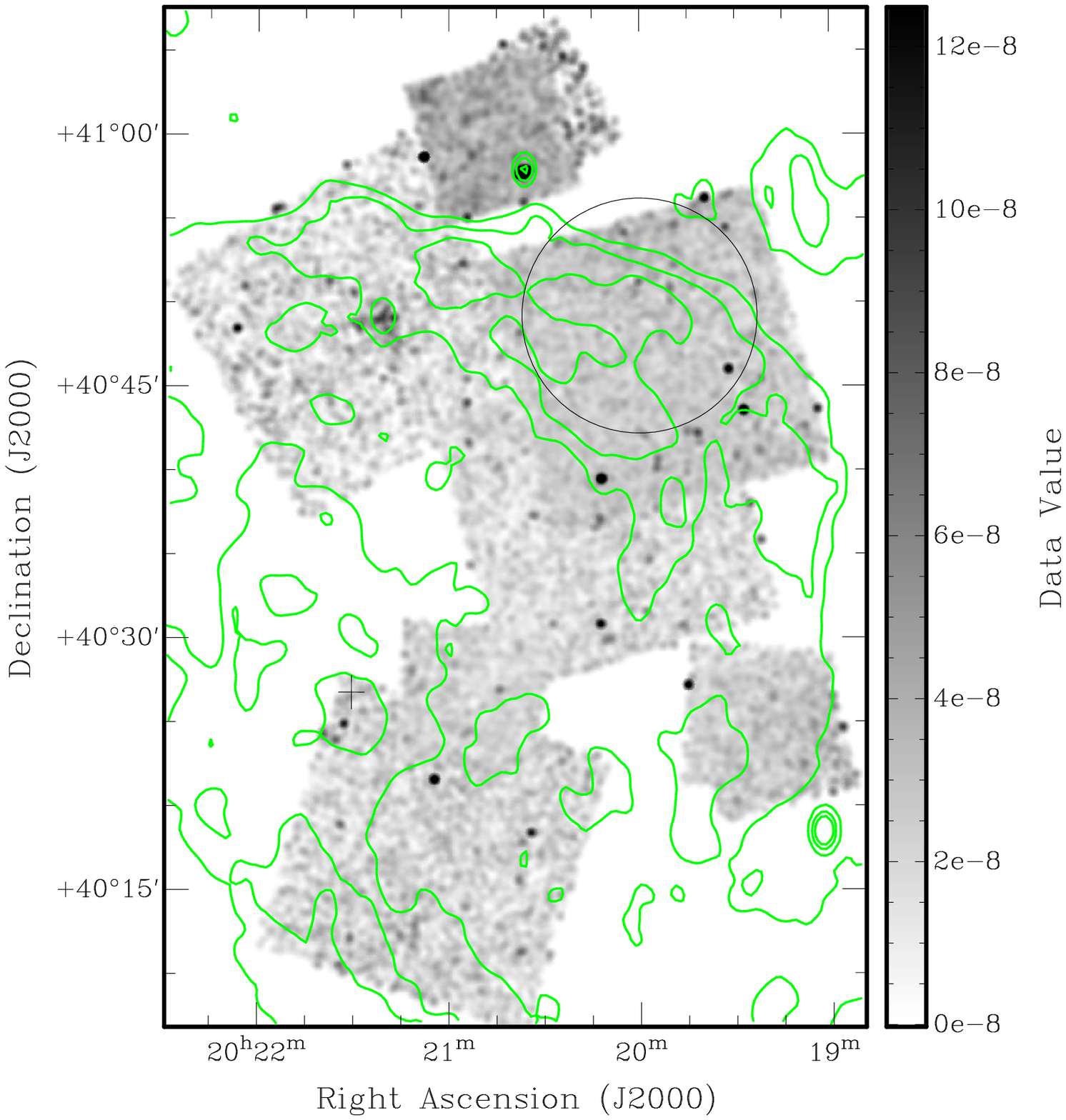}%{3-7_cts_ps.eps}
 \caption{ 3.0-7.0 keV \textit{Chandra} mosaic, overlaid with the 1420 MHz radio contours at T$_B$= 25, 32 and 42 K and smoothed with a 12 arcsecond Gaussian function. The black circle 
	indicates the error circle for VER J2019+407, the black plus indicates the position of 
	PSR J2021+4026.}
\end{figure*}

\subsection{X-ray spectrum analysis}

Because the \textit{Chandra} ACIS instrument has significantly superior spectral and spatial resolution to \textit{ROSAT}, 
we focus on the \textit{Chandra} spectra. 
Several regions were selected for X-ray spectrum extraction and analysis to understand the nature of the
X-ray emissions and where they originate, based on the derived column densities. 
Care was taken to avoid 
point sources for both the source and background regions. Thus each source and background
region was taken as the sum of spectra from a number of small rectangular areas (typically 4 to 12 boxes)
which avoided the visible point sources. 
Spectra were extracted using the procedure outlined in the CIAO Science threads and analysed using XSPEC.
The parameter uncertainties were determined using the 'error' function of XSPEC. 
The regions discussed here are: i) the center of G78.2+2.1
%extracted from observation ID 5533 and 
including the diffuse emission on the left side of the rectangular area
in the lower left part of Fig. 4 (centered on $\alpha$(J2000)=$\mathrm{20^h21^m00^s}$, $\delta$(J2000)=$40^{\circ}17^{\prime}00^{\prime\prime}$); 
ii) the diffuse emission
outside the northern edge of G78.2+2.1 at the upper right part of Fig. 4 (centered on $\alpha$(J2000)=$\mathrm{20^h19^m30^s}$, $\delta$(J2000)=$40^{\circ}53^{\prime}00^{\prime\prime}$); 
iii) the diffuse emission inside the northern edge of G78.2+2.1 at the upper right of
Fig. 4 (centered on  $\alpha$(J2000)=$\mathrm{20^h20^m00^s}$, $\delta$(J2000)=$40^{\circ}48^{\prime}00^{\prime\prime}$); 
iv) the diffuse emission inside the northern edge of G78.2+2.1 at the upper left of Fig. 3 (centered on  ($\alpha$(J2000)=$\mathrm{20^h21^m24^s}$, $\delta$(J2000)=$40^{\circ}52^{\prime}00^{\prime\prime}$); 
v) the extended emission associated
with the hard X-ray C1 source (centered on  $\alpha$(J2000)=$\mathrm{20^h21^m18^s}$, $\delta$(J2000)=$40^{\circ}49^{\prime}00^{\prime\prime}$, see Fig. 5); 
and vi) the 
point source emission associated with the hard X-ray C2 source (centered on  $\alpha$(J2000)=$\mathrm{20^h20^m12^s}$, $\delta$(J2000)=$40^{\circ}39^{\prime}00^{\prime\prime}$, see Fig. 5).
For all these regions, adjacent low surface brightness areas were used for background subtraction.

\subsubsection{Spectrum of G78.2+2.1 and the northern X-ray shell}

The X-ray spectrum from the diffuse emission at the center of G78.2+2.1 was modelled in XSPEC
using a non-equilibrium ionization plasma model (NEI) with a TBABS model for interstellar absorption. 
Solar abundance is assumed. An equilibrium ionization plasma model was found not to yield 
an adequate fit.
The resulting best-fit temperature is $\simeq$1 keV and column density is 
$N_H \simeq 9 \times 10^{21}cm^{-2}$. The best-fit parameters
and their 90\% error ranges are given in Table 3, along with parameters for the other spectra.
The column density derived from the HI absorption spectrum of G78.2+2.1
agrees, within errors, with that derived from the X-ray spectrum, confirming that the X-ray and radio emission
are both from G78.2+2.1.
Models of the emission from both the northern area of the remnant as well as the central region indicate that the plasma is not in ionization equilibrium. 

The spectrum from the diffuse emission just outside the edge of G78.2+2.1 was fit adequately with an
equilibrium ionization plasma (APEC) model with solar abundances. 
This spectrum has a much lower column density ($N_H \simeq 3 \times 10^{21}cm^{-2}$) than the central
spectrum and somewhat lower temperature. This shows than the northern X-ray emission outside G78.2+2.1
is from hot gas much closer to us than the SNR, and is not associated with G78.2+2.1.
The properties of this shell will be discussed in detail in Section 4.2.  

  \begin{figure*}
 \includegraphics[scale=0.33, angle=270]{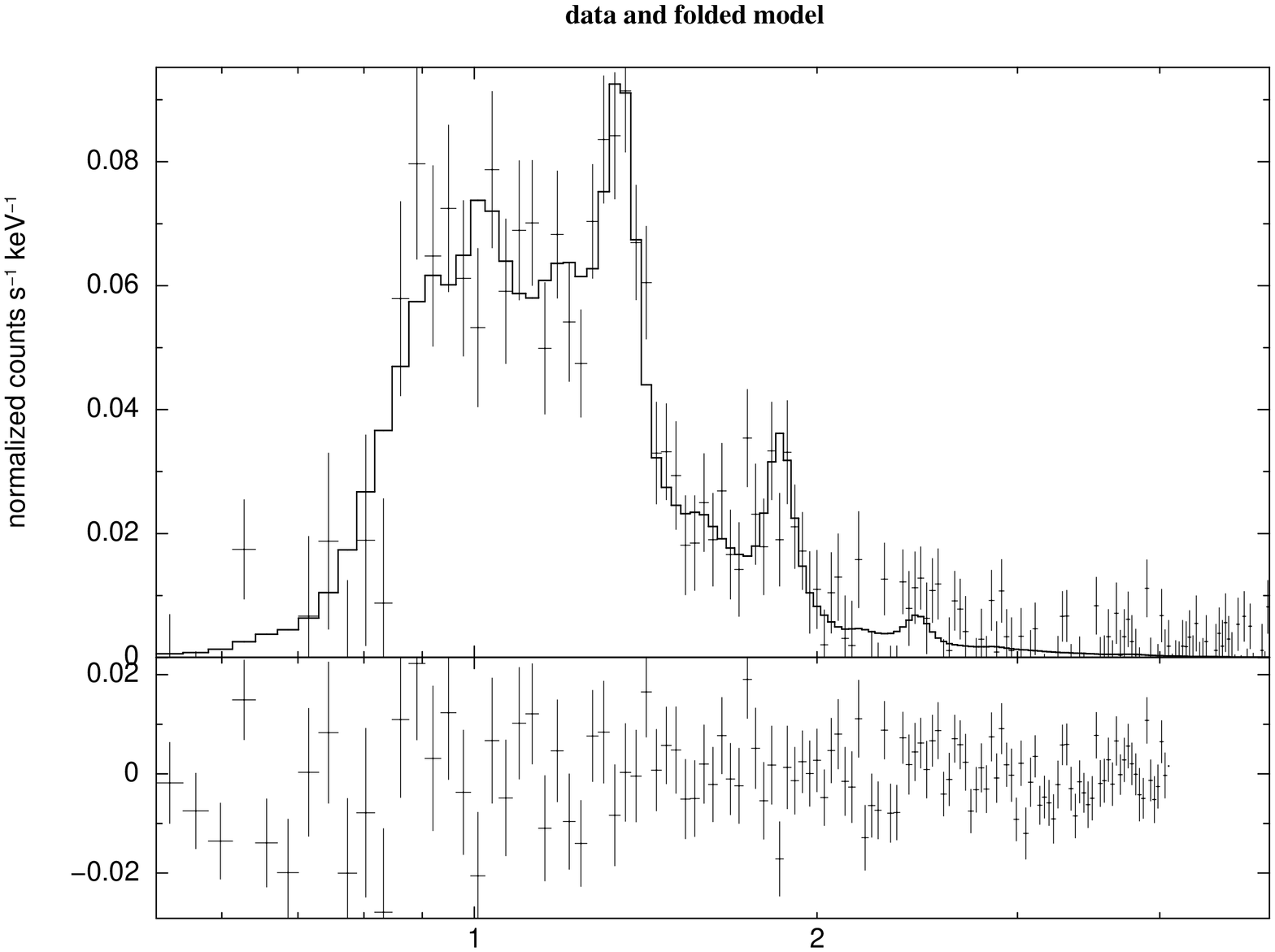} 
 \includegraphics[scale=0.33, angle=270]{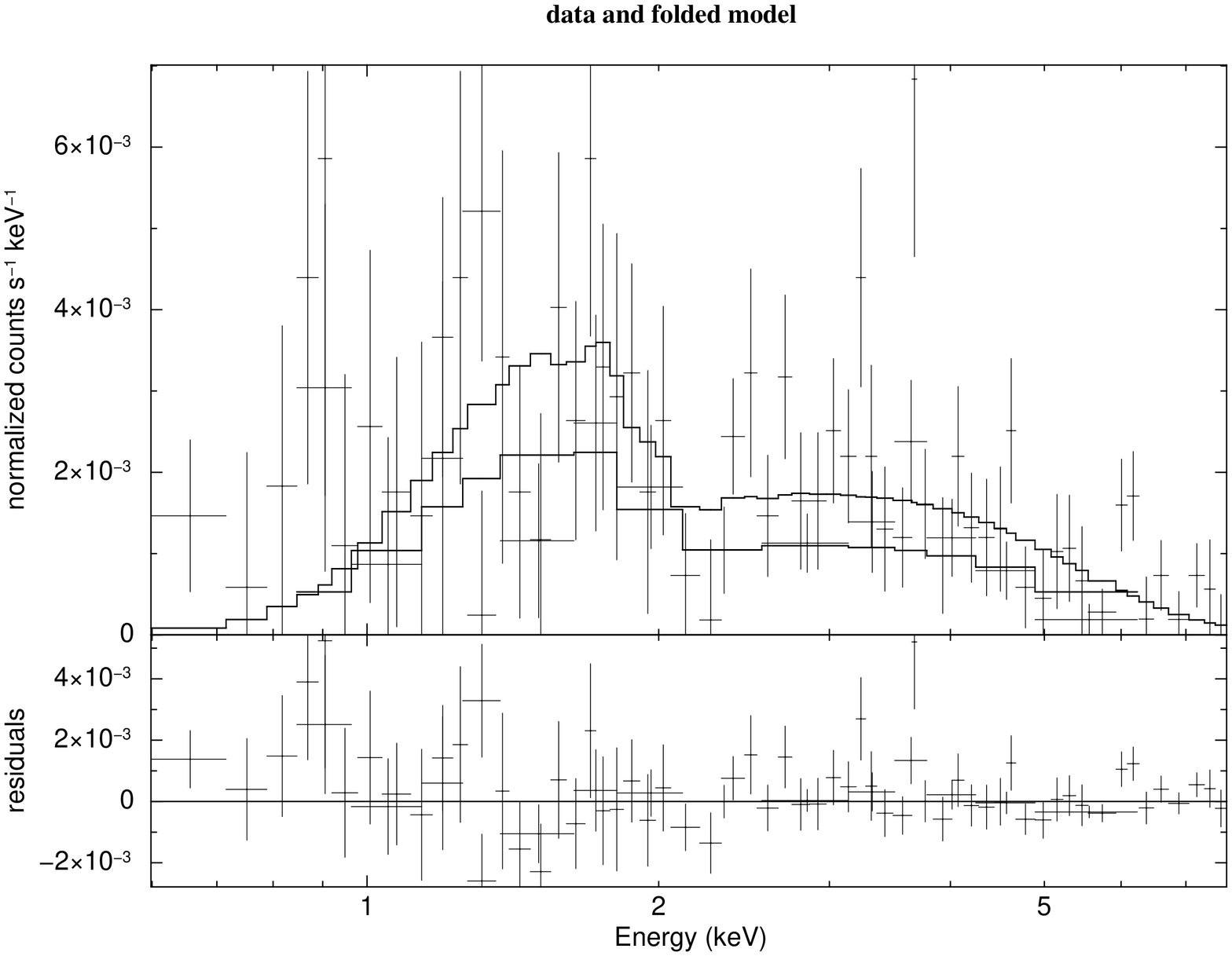}
 \caption{
 \textit{(left)} Spectrum of the SNR diffuse emission of the northern interior of G78.2+2.1, from I2 chip of observation 12676.
 \textit{(right)} Spectrum of the C2 point source from observations 12676 and 2826.}
\end{figure*}

The is some spatial overlap between G78.2+2.1 and the northern X-ray shell. 
The RASS image of Fig. 3a and especially the \textit{ROSAT} PSPC mosaic of Fig. 3b show that the overlap
along the northern boundary of G78.2+2.1 is significant.
Therefore we analyse the spectra from the northern edge G78.2+2.1 by allowing two spectral
components of different column densities in the spectral model.
We first extract a spectrum of the diffuse emission from the whole field covered by 
the I0, I1, I2 and I3 ACIS chips for observation 12676.
This field is the 5 arcmin by 5 arcmin rectangular region centered at 
($\alpha$(J2000)=$\mathrm{20^h20^m00^s}$, $\delta$(J2000)=$40^{\circ}47^{\prime}00^{\prime\prime}$) (see Fig. 4).  
The best-fit model for this data was a combination of an equilibrium ionization APEC model 
and non-equilibrium ionization (NEI) model. 
For this two component fit only, the XSPEC error command did not converge for 90\% errors but did converge for 1$\sigma$ errors so the 1$\sigma$ parameter ranges were used. 
The best fit parameter values are given in Table 3 (row 3 for APEC component, row 4 for NEI component). 
In addition to fitting the total diffuse emission, 
we separately fit spectra for the diffuse emission from the I2 and I3 ACIS chips 
(the I0 and I1 emission was too faint for separate fits that were useful). 
The results are shown in Table 3. We find the emission 
from the I2 chip of observation 12676 is consistent with that of the center of G78.2+2.1 and
the emission from the I3 chip is consistent with that found for the northern X-ray shell.
This is consistent with our result that the entire diffuse emission (I0, I1, I2 and I3) is a 
mixture of emission from both the G78.2+2.1 and the northern X-ray shell. 
As another check on the spectral fits, a joint spectral fit for the diffuse emission of this same
area was carried out using data extracted from both observations 12676 and 2826.
It was found that a two-component model was needed to fit the data, similar to the case for
the spectrum extracted for the diffuse emission
for observation 12676.

In summary, some of the emission from the northern part of G78.2+2.1 has
high column density and temperature, the same as the center of G78.2+2.1, 
and some of the emission is from the foreground northern X-ray
shell, with low column density and temperature.

Because the \textit{Chandra} observations do not cover the whole of G78.2+2.1, we do not have an estimate
of the total flux of the SNR nor its total emission measure from the \textit{Chandra} data. Thus we used
the \textit{ROSAT} pointed observations to extract a spectrum for whole area of G78.2+2.1. We find that
an APEC model fits the \textit{ROSAT} spectrum ($\chi^{2}$/dof=0.7) with  N$_H$ and kT consistent with the
Chandra values. The resulting APEC norm from the \textit{ROSAT} spectral fit is 0.11 with 90\% error range of 0.09 to 0.18.

\begin{table*}
\begin{minipage}{200mm}
\caption{\textit{Chandra} ACIS spectral fit parameters, SNR G78.2+2.1 and northern shell}
	\label{symbols}
	\begin{tabular}{@{}lccccc}
	\hline
	   & N$_H$(10$^{22}$cm$^{-2}$)  & kT(keV) & $n_et$(cm$^{-3}$s) & norm$^{(a)}$ & $\chi^{2}$/dof  \\
	\hline
center of SNR(5533) & 0.92(0.75-1.11) & 1.05(0.59-2.73) & 0.33(0.17-1.2)$\times10^{11}$& 1.3(0.0-3.5)$\times10^{-3}$ &1.12 (199/177)  \\
northern shell (12676) & 0.30(0.19-0.39) & 0.69(0.64-0.75) & - & 0.19(1.4-2.4)$\times10^{-3}$ & 0.98 (159/162)  \\
diffuse emission (12676)$^{(b)}$ & 0.39(0.28-0.39) & 0.75(0.71-0.75) & -& 3.7(1.9-4.5)$\times10^{-4}$ &  \\
								&1.1(0.91-1.1) & 1.2(1.0-2.0) & 3.0(2.0-3.6)$\times10^{10}$ & 1.5(0.96-1.7)$\times10^{-3}$ &1.11 (332/298) \\
SNR	diffuse emission (I2 of 12676) & 0.99(0.90-1.13) & 0.75(0.57-1.16) & 0.81(0.36-5.7)$\times10^{11}$ & 1.3(0.71-2.4)$\times10^{-3}$ & 1.27 (189/149)\\
N.shell diffuse emission (I3 of 12676) & 0.23(0.13-0.30) & 0.57(0.52-0.63) & - & 1.5(1.1-1.9)$\times10^{-3}$ & 0.91 (99/109) \\
	\hline
	\end{tabular}
  $(a)$norm=$10^{14}/(4\pi D^2) \int n_e N_H dV$ \\
  $(b)$Fit of diffuse emission from all ACIS-I chips,  
1$\sigma$ error values used. 
\end{minipage}
\end{table*}

\subsubsection{Uchiyama hard sources C1 and C2}
With the improvement in spatial and spectral resolution of \textit{Chandra} over ASCA, 
we re-examine the hard sources C1 and C2 in G78.2+2.1 found by Uchiyama et al (2002). 
We see clear evidence that C1 is an extended source in both 0.5-3 keV and 3.0-7.0 keV \textit{Chandra} images 
(see Figs. 4 and 5 at the C1 position of $\alpha$(J2000)=$\mathrm{20^h21^m24^s}$, $\delta$(J2000)=$40^{\circ}49^{\prime}00^{\prime\prime}$). 
We extract the C1 spectrum from a circle of radius 2' centered on C1. 
There is significant radio emission associated with C1. This is indicated by the 42K contour line 
in the 3-7 keV image. This is also verified in the 1420 MHz image (Fig. 1), where it appears as a
bright compact source at l=78.52$^\circ$, b=+2.33$^\circ$, 
as it is not resolved in the radio image.
The spectrum is analysed using a combination APEC plus power-law model, with TBABS model 
for interstellar absorption. As seen in Table 4, the column density is consistent with C1 being an extragalactic background source.  

The C2 region is defined by Uchiyama et al (2002) is centered on ($\alpha$(J2000)=$\mathrm{20^h20^m06^s}$, $\delta$(J2000)=$40^{\circ}45^{\prime}00^{\prime\prime}$) 
and 6' in extent.
From the soft and hard \textit{Chandra} X-ray images, we see that the emission from the C2 region is dominated by
a hard X-ray source at the center of the region, which we also designate C2. 
In Table 4 we show the spectral fit for the point source, which is a simultaneous fit of spectra extracted from datasets 12676 and 2826, as both include the C2 region.
We also fit the spectra for C2 from 12676 and 2826 separately, with results that are consistent with
the values presented in Table 4, but with larger errors. 
The best-fit spectrum is a powerlaw with a column density consistent with that for G78.2+2.1.
However, due to the low count rate and resulting large errors on $N_H$, it is not possible to 
say definitively whether or not C2 is associated with G78.2+2.1. 

\begin{table*}
\begin{minipage}{200mm}
\caption{\textit{Chandra} ACIS spectral fit parameters$^{(a)}$: Hard Sources C1 and C2}
	\label{symbols}
	\begin{tabular}{@{}lcccccc}
	\hline
	   & N$_H$(10$^{22}$cm$^{-2}$)  & kT(keV) & norm(plasma) & photon index & norm(powerlaw) & $\chi^{2}$/dof  \\
	\hline
	C1(2827) & 1.5(1.2-1.9) & 0.41(0.30-0.52) & 1.1(0.5-3.7)$\times10^{-3}$ & 0.6(0.1-1.0) & 
	2.7(1.5-5.2)$\times10^{-5}$ & 1.0 (31/31) \\
	C2(point source)$^{(b)}$ & 0.69(0.025-1.3) &-& - & 1.0(0.72-1.4) & 2.3(1.0-4.2)$\times10^{-5}$  & 1.0 (70/70) \\
	\hline
	\end{tabular}
	\\
  $(a)$ 1$\sigma$ error values used. \\
	${(b)}$ Simultaneous fit from datasets 12676 and 2826. \\
\end{minipage}

\end{table*}

\begin{figure*}	
	\includegraphics[scale=0.32, angle=270]{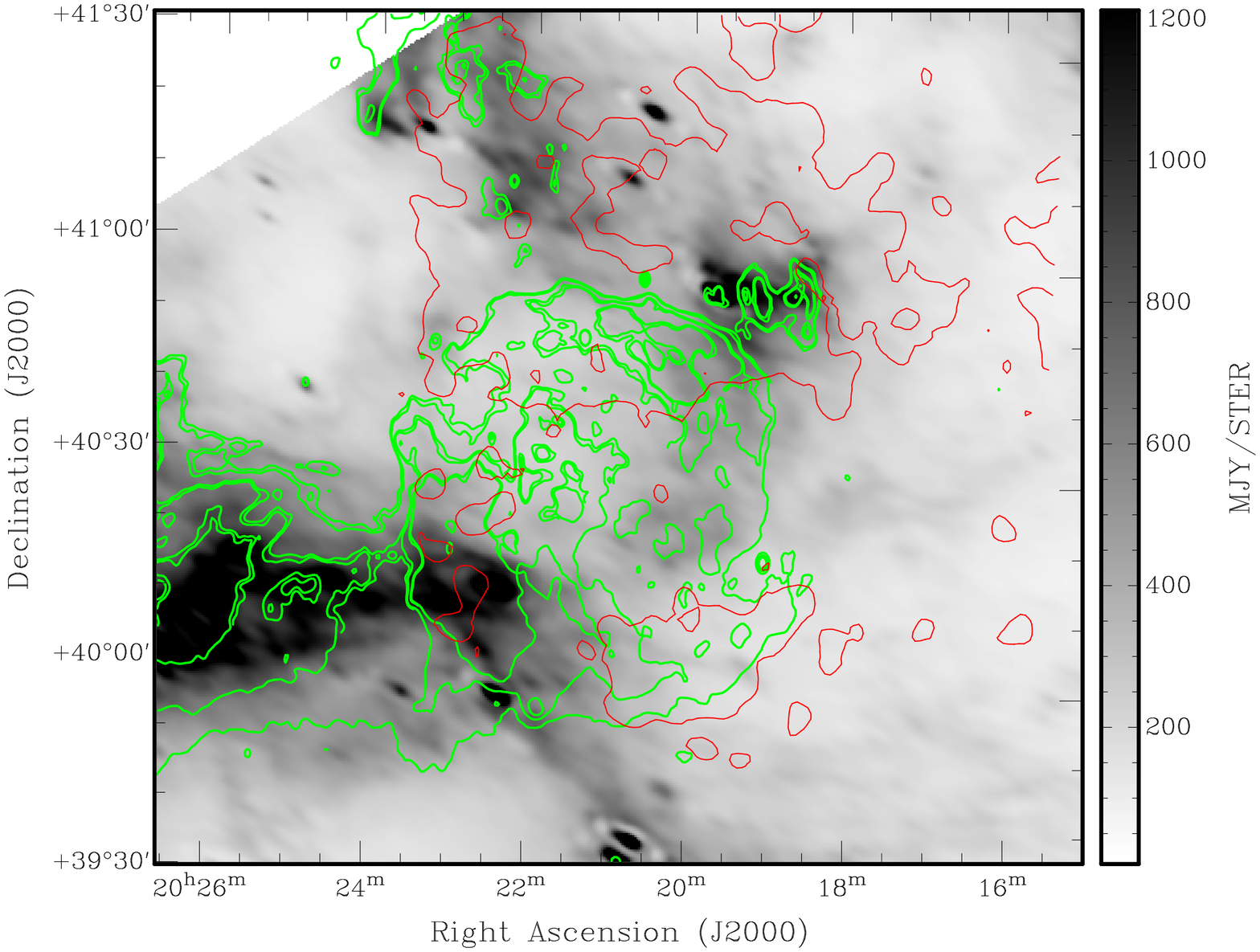}
	\includegraphics[scale=0.32, angle=270]{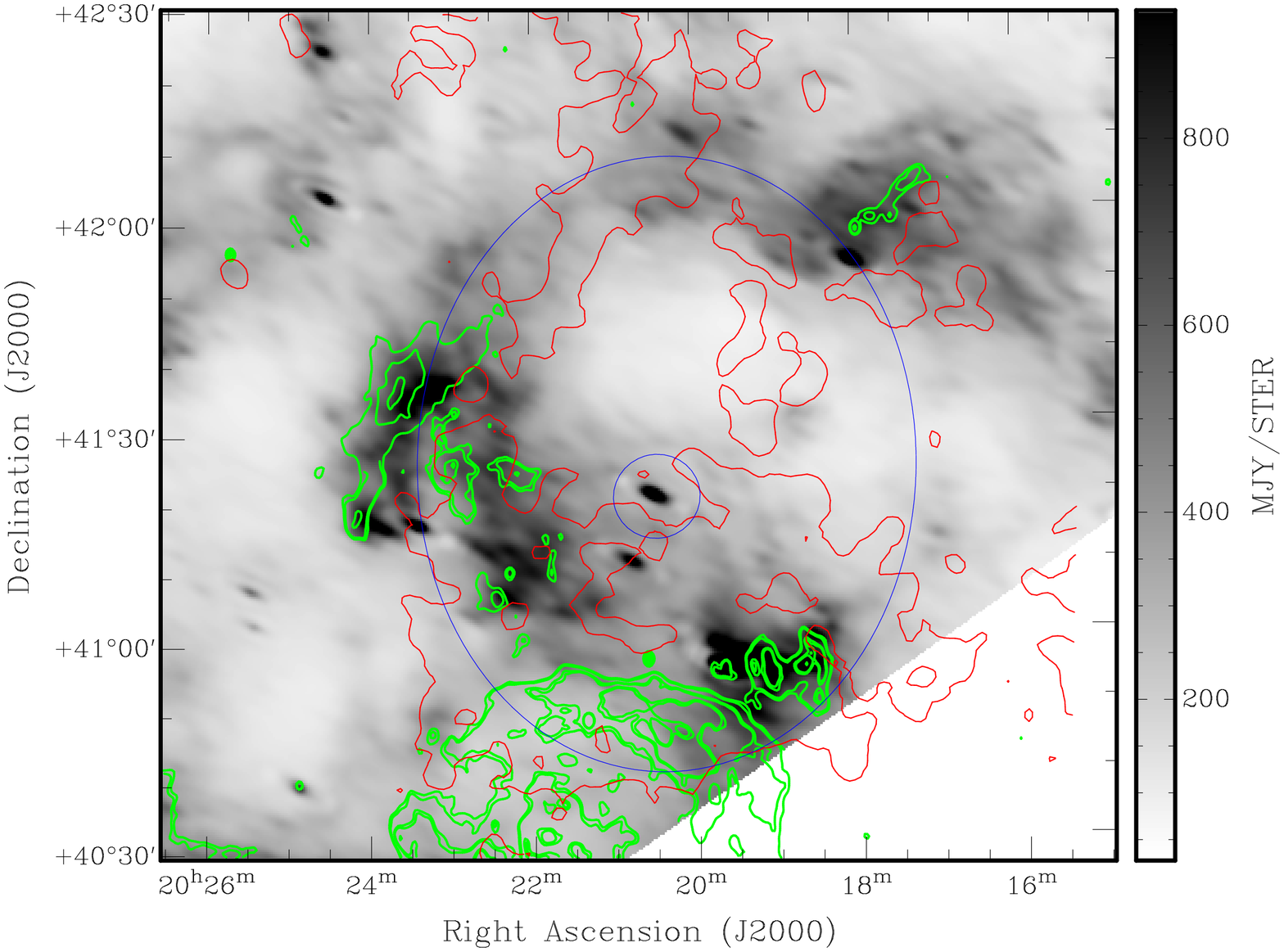}
	\caption{ 
	 IRAS 60 micron image of SNR G78.2\textit{(left)} and of the northern X-ray shell\textit{(right)}, with contours from the RASS survey image (red) and 1420MHz radio image (green). The blue oval indicates
	 the approximate location of the shell, and the small blue circle indicates the early-type star V1685 Cyg.}
\end{figure*}

\section{Discussion}

\subsection{The SNR G78.2+2.1}

\subsubsection{HI and the distance to G78.2+2.1}

We produce the first quantitative absorption spectra for SNR G78.1+2.1 and the $\gamma$ Cygni nebula.
Higgs et al.(1977) present an emission plus fractional absorption 
profile which shows similarity to our absorption spectrum. However, quantitative 
comparison with our results is not possible because they do not explain how their 
combined emission and absorption profile is calculated, nor do they give 
the location of their extraction regions.

The most negative velocity of absorption is -8 km/s, consistently for the different HI absorption
spectra that were extracted for G78.2+2.1. 
To interpret this in terms of distance is not simple.
%Allowing for some velocity dispersion in the absorbing gas we estimate a systematic velocity for the SNR of -4 to -8 km/s. 
%Using a galacto-centric radius for the Sun of R$_0$=8 kpc and a circular rotation curve 
%for HI in the region of R=7 to 9 kpc with V=220 km/s, the above systematic velocity
% gives a distance of 3.9 to 4.5 kpc. 
As noted by Gottschalk et al.(2012), radial velocities in the Cygnus X region,
of which G78.2+2.1 is part, do not follow a circular rotation model. Rather
the molecular gas is in three layers (see introduction above). Since all three layers 
show up in absorption in the HI spectrum of G78.2+2.1, the SNR must be either in or
behind the furthest of the layers, layer 2, which is at a distance of 
1.5-2.5 kpc (Gottschalk et al. 2012), different from its distance inferred using a circular rotation curve model. 
Thus 1.5-2.5 kpc lower distance limit of layer 2 is taken as the lower limit to the distance to the SNR. 
%Regarding an upper distance limit, we note again that velocities in the Cygnus X
%region are not consistent with a Galactic circular rotation model. 
We note that 
Gottschalk et al. (2012) show relative absorption profiles for sources in Cygnus X for
a region 3-4 degrees away from G78.2+2.1. 
Two of the sources have measured parallax distances: 1.3 kpc for W75N and 1.5 kpc for DR21 (Rygl et al. 2012).
These two sources have different maximum velocities of
absorption: $\sim$+3 km/s for W75N and $\sim$-20 km/s for DR21 (Gottschalk et al. 2012) . 
Thus the absorption spectrum is not of use to obtain any better distance limit than the lower limit
mentioned above.

Ladouceur \& Pineault (2008) discusses HI emission in the region.
Their Figs. 19, 21 and 23 shows HI emission averaged over channels for
velocities of +3 to +18 km/s, +3 to +9 km/s and -19 to -11 km/s, respectively. 
%It is not  clear whether the edge of the HI which agrees with the edge of the 
%continuum emission is entirely caused by HI absorption or not. 
For +3 to +18 km/s they claim that the low intensity features are absorption,
but for +3 to +9 km/s they claim that the same low intensity features are caused
by a shell surrounding the SNR, despite the overlapping velocity range.
We believe a more consistent explanation is that all features are caused by absorption.
Their Fig.24 shows the CO emission averaged from +3 to +10 km/s: we see no clear 
association with SNR G78.2+2.1. 
A sharp edge is seen to the CO emission near
l=79 degrees, from b=0 to 1.5 degrees, but this is more than 1 degree south of the SNR.
Ladouceur \& Pineault (2008) also propose four scenarios for the location of HI of
different velocities around or in front of SNR G78.2+2.1. However none of the  
scenarios is consistent with the recent findings of Gottschalk et al (2012), which
show clear evidence for three HI layers, with
the 0 to +8 km/s layer nearest to us and well in front of the Cygnus X region and
the G78.2+2.1. 

In section 3.1 we noted the detailed agreement between the absorption spectra for the SNR and 
the $\gamma$ Cygni nebula. 
We conclude that they are probably physically associated and at the same distance.
The distance to the $\gamma$ Cygni nebula is 1.7 to 2.6 kpc (Baars et al., 1978).
This is consistent with the lower distance limit of 1.5 kpc, derived above.
%from the SNR and the $\gamma$ Cygni nebula both behind Layer 2 of Gottschalk et al (2012).
Recently, Rygl et al (2012) have determined accurate distances for 
five massive star forming regions in the Cygnus X region. Four of the five have 
distances of 1.3 to 1.5 kpc, yielding the best distance yet for the Cyg OB2 
association, at the center of Cygnus X. The fifth region is AFGL 2591, which 
at $\alpha$(J2000)=$\mathrm{20^h29^m24.8^s}$, $\delta$(J2000)=$40^{\circ}11^{\prime}19.6^{\prime\prime}$ is located much closer to G78.2+2.1
than the others. It has a distance of 3.33$\pm$0.11 kpc. These are similar to the distance
to the  $\gamma$ Cygni nebula, but there is not yet a better distance determination to the latter.

In summary, we take the distance 1.7 to 2.6 kpc of the  $\gamma$ Cygni nebula as the
probable distance for G78.2+2.1. 

\subsubsection{X-ray properties}

The new X-ray results allow us to determine the SNR properties better than previously
done.
First we apply a Sedov model (e.g. Cox, 1972) to G78.2+2.1 with distance range 1.7-2.6 kpc. 
The SNR angular radius is 30 arcmin, so with distance $D=D_{2}\times2$ kpc the SNR radius 
is $R_s$=17$D_{2}$pc. 
The shock temperature of 0.75 keV is taken from the \textit{Chandra} spectra of the SNR diffuse emission,
which has a 90\% uncertainty range of 0.6-1.2 keV (see Table 3). 
This yields a shock velocity of 860 km/s (90\% error 700-1100 km/s).
Including a distance range of 1.7-2.6 kpc, the Sedov age is 8000 (6800,10000) yr and 
the parameter $\epsilon_0 /n_0$ is 7.2 (4.4,16), with upper and lower limits in brackets. 
Here $\epsilon_0$ is the blast energy in units of 0.75$\times10^{51}$erg. 
The norm from the \textit{ROSAT} APEC spectral fit for the whole of G78.2+2.1
gives an emitting plasma density of % $n=0.37/(f_{0}D_3)$
$n=0.56/(f_{0}D_2)$ where $f_{o}$ is the filling factor in 
units of 0.05. The fiducial filling factor of 0.05 is obtained using the density distribution for a
Sedov remnant, e.g. as given by Heiles (1964).
The norm of the  \textit{ROSAT} spectral fit also yields a mass of the X-ray emitting plasma
of $\sim$ 50 M$_{\odot}$.
We checked, using the \textit{Chandra} images which resolve point sources, that the contribution of
point sources to the \textit{ROSAT} norm is small ($\sim <$20\%, so that the \textit{ROSAT} norm is a good approximation
to the norm of the diffuse emission from the SNR.
The norm then gives an estimate for the pre-shock density, using a compression ratio 
of 4 for a strong shock, of $n_0=n/4 \simeq 0.14$.
This in turn gives an explosion energy of 0.8(0.5,1.7)$\times10^{51}$erg.
%note epsilon in units of 0.75e51 erg
Finally we test whether the SNR is expected to be old enough to be past the Sedov (adiabatic) phase
by calculating $t_{sag}$, the time at which radiation losses cause the temperature to sag just 
behind the forward shock (Cox, 1972). With the above parameters we find
$t_{sag}$=4.7(4.3,5.4) $\times 10^4 \times (10^{22}\Lambda)^{-5/11}$ yr, with 
$\Lambda \sim 10^{-22}$ the radiative cooling coefficient. 
Thus the Sedov age is significantly below the age at which cooling sets in,
and the Sedov approximation should be a good approximation.

Our X-ray spectrum analysis is consistent with the ASCA results of Uchiyama et al. 
(2002), who found diffuse plasma temperatures of 0.6-0.8 keV, but with the spatial
resolution and spectra of the \textit{Chandra} data, we can separate the emission from 
the SNR from that of compact sources and from that of the foreground X-ray shell. 
The Mavromatakis (2003) estimate of the column density of $6\times10^{21}$cm$^{-2}$
 agrees within errors with that we obtain from the HI 
absorption spectrum and from the \textit{Chandra} spectral analysis.

For the compact sources C1 and C2, the higher resolution \textit{Chandra} observations allow us to clearly
separate the emission from the compact sources from the surrounding X-ray diffuse emission.  
C1 is an extended hard X-ray source with column density indicating that it is
an extragalactic background source. 
C2 is a point hard X-ray source with a pure power-law spectrum and
has column density consistent with the SNR G78.2+2.1, and may be associated with the SNR.
The 0.5-3 keV Chandra image (Fig. 4) also shows the location of the $\gamma$-ray pulsar PSR J2021+4026, 
which like C2 has a column density consistent with that of the SNR, and the error circle of the TeV source VER J2019+407.
C2 is outside the 95\% error circle of VER J2019+407, so is probably not associated with it.
VER J2019+407 is extended (Weinstein et al. 2011) and coincides with the rim G78.2+2.1 in
diffuse radio and X-ray emission, so is likely associated with the SNR.
Inconclusive evidence indicates PSR J2021+4026 may be associated with G78.2 (Trepl et al, 2010).
It seems clear that further observations are needed to settle the relationship between the 
three sources (PSR J2021+4026, VER J2019+407 and C2) and the SNR.
 
\subsection{The northern X-ray shell}

The northern X-ray shell, seen clearly in the RASS image of Fig. 3(a), was found to have a low column
density from a \textit{Chandra} spectrum taken near its southern edge, but outside of the SNR G78.2+2.1.
We searched for counterparts to the shell and see diffuse infrared emission coincident with
the shell using IRAS All-Sky-Survey images. This is shown in Fig. 7, with the left panel showing the
60 micron image around G78.2+2.1 and the right panel showing the 60 micron image around the northern
shell. 
We also found an early type star near the shell's center.
The star is V 1685 Cyg (=HIP 100289), with coordinates: ($\alpha$(J2000)=$\mathrm{20^h20^m28^s}$, $\delta$(J2000)=$41^{\circ}21^{\prime}52^{\prime\prime}$),
or Galactic l=78.8719d,b=+02.7699d. From SIMBAD, the apparent V magnitude and B-V color are:
V=10.88, B-V=0.57. The spectral type is B3 and its distance is 980 pc (Manoj et al., 2006),
which is consistent with the X-ray column density.
A B3 main sequence star has absolute magnitude $M_V=-1.6$ and intrinsic color $(B-V)_0=-0.20$ 
(Binney and Merrifield, 1998).
The observed V and B-V, then yield an extinction estimate of $A_V=2.39$ (using $A_V=3.1 E_{B-V}$, 
and a distance based on its extinction-corrected apparent magnitude of 1.0 kpc, in agreement
with the published value.

To estimate the physical properties of the shell, we extract the spectrum of the entire
shell using the RASS survey data. A background spectrum is extracted 
from an adjacent area just east of the shell. The spectrum is fit using an APEC model and
yields the following parameters (with 90\% errors in brackets): 
$N_H=4(3,6.5)\times10^{21}$cm$^{-2}$, kT$=0.24(0.19,0.31)$keV, norm=1(0.3,1.7).
The norm in XSPEC is $(10^{14}/D^2) \int n_e N_H dV$, which then yields a
density of 0.11 f$^{-1/2}$ cm$^{-3}$, where f is the filling factor.
We tested that a Sedov SNR model is inconsistent with the northern shell, because it results in
an unreasonably small explosion energy.
  
Binney and Merrifield, 1998, give standard parameters for a B3 main sequence star:
M= 7.6 M$_\odot$, L=$10^{3.28}$L$_\odot$, R=4.8 R$_\odot$, T$_{eff}$=18700K.
These can be used to estimate what the properties of a stellar wind bubble should be,
where we use the stellar wind bubble model of Kwok (2007). 
To estimate the stellar wind power for a star with this T$_{eff}$, L and M, we use the 
formula from Nieuwenhuijzen \& de Jager (1990), with wind velocity $\sim$ 2 times the
escape velocity. %$V_{wind} \sim 2 V_{esc}$. 
This yields a stellar wind power of $\sim 4\times 10^{32}$erg s$^{-1}$. Then the observed
angular radius and distance give the bubble radius of 8.5 pc. Applying the wind bubble
model then gives an age for the bubble of $\simeq 2\times 10^6$yr.

\section{Conclusions}  

We carry out a new study of the supernova remnant G78.2+2.1 using 
Chandra and \textit{ROSAT} X-ray observations and CGPS radio continuum and HI line observations.
We find that HI absorption spectra of G78.2+2.1 and the $\gamma$ Cygni nebula are nearly identical, so we associate them at the same distance of 1.7-2.6 kpc (Baars et al. 1986).
This is also consistent with the lower limit of 1.5 kpc determined by absorption from gas layer 2 
in the Cygnus X region (Gottschalk et al, 2012, and Rygl et al, 2012).
The \textit{ROSAT} All-Sky-Survey image shows a bright X-ray shell to the north of G78.2+2.1, and
overlapping G78.2+2.1 on the SNR's northern boundary.
We make a new high resolution \textit{ROSAT} PSPC mosaic of G78.2+2.1 showing for the first time the full
extent and structure of the SNR G78.2+2.1 in X-rays.
Chandra ACIS archival images of G78.2+2.1 are used to construct high-sensitivity, high 
resolution X-ray images in soft and hard X-ray bands.
Chandra ACIS  X-ray spectra of the center of G78.2+2.1 
yield a column density of $N_H \simeq 9 \times 10^{21}$cm$^{-2}$, 
in agreement with the value from the HI absorption spectrum.
A Sedov model is applied to G78.2+2.1 yielding an age of 6800-10000 yr and an explosion energy
of 0.5-1.7$\times10^{51}$erg.
A spectrum from the northern X-ray shell gives a
column density of $N_H \simeq 3 \times 10^{21}$cm$^{-2}$, showing that
it is much closer to us than G78.2+2.1. 
%Furthermore, spectral analysis for the northern edge of
%G78.2+2.1, which overlaps the northern X-ray shell, shows the emission is a mixture of low column
%density emission with temperature the same as for the northern shell, and high column density
%emission with with temperature the same as for the center of G78.2+2.1.

Two prominent hard X-ray sources C1 and C2 were identified in ASCA observations of G78.2+2.1
(Uchiyama et al. 2002). We use
the \textit{Chandra} ACIS observations, with their superior spatial and spectral resolution, to show
that C1 is probably an extragalactic extended source, and that C2 is a point source with a 
power-law spectrum with the same column density as G78.2+2.1.
For the northern X-ray shell, %is seen clearly in the IRAS All Sky Survey. We have found 
an early-type
(B3 main sequence) star V1685 Cyg, at distance 980 pc, is found at the center of the shell. 
This star is identified as the star 
responsible for creating and heating the shell via its stellar wind.

\section{Acknowledgements}
This work supported by the Natural Sciences and Engineering Research Council of Canada.  The Dominion Radio Astrophysical Observatory is operated as a national facility by the National Research Council of Canada. We thank the referees for important suggestions which lead to improvements in the paper.

\end{document}